\documentstyle[12pt,epsf]{article}
\newcommand{\gev}{\, \mbox{GeV}}

\newcommand{\beq}{\begin{equation}}
\newcommand{\eeq}{\end{equation}}
\newcommand{\Frac}[2]{\frac{\displaystyle #1}{\displaystyle #2}}

\begin{document}
\thispagestyle{empty}
\begin{titlepage}
\begin{center}
\vspace*{-1cm}
\hfill IFIC/00$-$19 \\
\hfill FTUV/00$-$0331 \\
\hfill TAN-FNT-00$-$02
\vspace*{2cm} \\
{\Large \bf The hadronic off--shell width of meson resonances}
\vspace*{1.6cm} \\
{ \sc D.\ G\'omez Dumm$^a$, A.\ Pich$^b$} and {\sc J.
Portol\'es$^b$}
\vspace*{0.7cm} \\
$^a$ Departamento de F{\'{\i}}sica, Comisi\'on Nacional de
Energ{\'{\i}}a At\'omica, \\
Av.\ Libertador 8250, (1429) Buenos Aires, Argentina \\ \hfill \\

${}^b$ Departament de F\'{\i}sica Te\`orica, IFIC,
CSIC -- Universitat de Val\`encia \\
Edifici d'Instituts d'Investigaci\'o, Apt.\ Correus
2085, E--46071 Val\`encia, Spain \\
\vspace*{1.3cm}
\begin{abstract}
Within the resonance chiral effective theory we study the dressed
propagators of the spin--1 fields, as arise from a Dyson--Schwinger
resummation perturbatively constructed from loop diagrams with
absorptive contributions in the s--channel. We apply the procedure to
the vector pion form factor and the elastic $\pi \pi$ scattering to
obtain the off--shell width of the $\rho^0$ meson.
We adopt a definition of the off--shell width
of spin--1 meson resonances that satisfies the requirements of
analyticity, unitarity, chiral symmetry and asymptotic behaviour
ruled by QCD.
To fulfil these constraints the resummation procedure
cannot consist only of self--energy diagrams. Our width
definition is shown to be independent
of the formulation used to describe the spin--1 meson resonances.
\end{abstract}

\vfill
PACS numbers: 12.38.Aw, 12.38.Cy, 12.39.Fe, 12.40.Vv
\end{center}
\end{titlepage}
\newpage
\pagestyle{plain}
\pagenumbering{arabic}

\section{Introduction}
\hspace*{0.5cm} The evaluation of hadronic current matrix elements in the
low--energy regime is a long--standing problem of particle physics that has
been addressed using many different tools. The common lore amounts to
obtain momentum--dependent form factors that carry the dynamical
content of the interaction, though a rigorous determination in the
framework of a lagrangian formalism like the Standard Model has not yet
been achieved. From a field theoretical point of view, the goal is to
evaluate the Green's functions from the quantum action functional
but, in practice, this is a poorly known procedure overcome by many
incertitudes that arise because of hadronization and analytic continuation.
As a consequence we rely in the construction of form factors from
guiding principles such as analyticity, symmetries of the Standard Model, or
model--dependent assumptions of dynamical content (vector meson
dominance, factorization, duality and so forth). A more phenomenological
approach consists in fitting ad hoc parameterizations with experimental
data.
\par
Hadronic low--energy phenomenology far below the resonance region
($E \ll M_{\rho}$) is successfully described in the framework of
Chiral Perturbation Theory ($\chi PT$) \cite{WE79,GH84,GH85},
the effective action of
Quantum Chromodynamics (QCD) at low energies. However a similar tool
in the resonance region, typically $E \sim 1 \, \gev$, is still
lacking, and has become a focus of interest in the last years.
Data on $\pi \pi \rightarrow \pi \pi$, the vector pion form factor,
hadronic $\tau$ decays and other processes have prompted the
activity on the theoretical side.
\par
Following the phenomenological lagrangian ideas of Ref. \cite{CO69},
the inclusion of meson resonances in an effective theory was addressed
in Ref. \cite{EG89} by introducing the leading resonance chiral
effective theory
of QCD. It has to be emphasized that this is a model--independent
framework that provides \cite{EG89a}, upon integration
of the spin--1 meson degrees of freedom, the same ${\cal O}(p^4)$
$\chi PT$ lagrangian independently of the definition of the spin-1
fields, a non--trivial feature.
\par
The problem we wish to study comes
from the obvious fact that, in the resonance region, the width of the
resonances cannot be neglected and one needs to regularize the poles
in a proper way.
Here we address the construction of the off--shell width of
meson resonances from the resonance chiral effective theory of
QCD. We will focus, owing to its relevance
and simplicity, in the off--shell width of the $\rho^0$. As it is well
known the $\rho$ meson plays a crucial role in the dynamics of hadron
processes at low energies and therefore its properties, like mass and
width, have been thoroughly studied in many different frameworks
\cite{listarho,PT97} usually related with the pion form factor.
Notwithstanding most of these studies rely on modelizations of the
interaction that include assumptions not justified from the Standard
Model.
\par
In Ref.\cite{PT97} it was pointed out that one could consider the evaluation
of the off--shell $\rho$ width from the effective theory of QCD at low
energies that includes the resonances and Goldstone bosons explicitly. Here
we carry out in detail such a procedure.
We study two physical observables: the vector pion form
factor and the $\pi^+ \pi^- \rightarrow \pi^+ \pi^-$ amplitude with
$J=1$ in the s--channel.
We construct a Dyson--Schwinger--like equation and we
show that, in both observables, it gives the same imaginary part for
the $\rho^0$ pole.
\par
When considering off--shell processes one
might worry about the fact that different redefinitions of the fields
give different results. We have shown that the result of our procedure
does not depend on the definition of the $\rho^0$ field.
One of the conclusions of Ref.\cite{EG89a} was that, as different
redefinitions of the spin--1 fields give different lagrangians, interaction
vertices are dependent on the formulation and, therefore,
the Feynman diagrams contributing to a process are formulation--dependent.
We will have to take this fact into account when considering which
diagrams have to be accounted for in the
Dyson--Schwinger series. As a result this
is not going to be reduced to evaluate resonance self--energies only.
\par
We will proceed then by proposing a definition of the off--shell
width of the vector meson resonances as the imaginary part of the pole
generated through the two--point vector current correlator, where only
those diagrams with an absorptive part in the s--channel are included.
This definition is shown to satisfy the crucial requirements of analyticity,
unitarity, chiral symmetry and asymptotic behaviour ruled by QCD.
Again the result is shown to be independent of the spin--1 field
formulation.
\par
In Section 2 we study the resummation that regulates the pole of
the $\rho^0$ in the vector pion form factor and the elastic
scattering process $\pi^+ \pi^- \rightarrow \pi^+ \pi^-$. Then in
Section 3 we propose a general definition of the off--shell width
of resonances. Section 4 is devoted to provide the rationale for
the independence of our results from the spin--1
meson formulation. Our conclusions are pointed out in Section 5.

\section{The pole of resonances in physical observables}
\hspace*{0.5cm} The position of the pole of the bare propagator for
stable particles gets shifted when interactions are switched on. In
the usual perturbative treatment of the interaction, a pole has
to be achieved through a resummation procedure of higher orders. The
well known solution of the Dyson--Schwinger equation for the dressed
propagators,
obtained through the evaluation of self--energy Feynman diagrams, hides
the fact that the definition of the resummation, that is, which
are the contributions and how one has to proceed,
is not free of ambiguities. These are
lessen if one needs to impose more restrictions on the result, like
gauge invariance
(see \cite{PP96} and references therein). However, hadron
processes at low energies are described by an effective action where
colour $SU(3)$ gauge invariance is not explicit. At the $1 \, \gev$
scale we are driven by two
all--important features: chiral symmetry and the asymptotic behaviour
of form factors ruled by QCD. Obviously basic principles like
analyticity and unitarity must also be satisfied.
\par
Chiral symmetry has a long history as a powerful tool to describe
low--energy hadrodynamics \cite{PI95}. One of its main
aspects is that it is a spontaneously broken symmetry that requires
the existence of Goldstone bosons to be identified with the octet of
lightest pseudoscalars.
Chiral Perturbation Theory is the
effective action of low--energy QCD \cite{WE79,GH84,GH85} that, in the
$SU(3)_L \otimes SU(3)_R$ version, involves only the octet of pions,
kaons and $\eta$ mesons and describes strong interactions at
$E \ll M_{\rho}$. At the $1 \gev$ energy region the inclusion of the
lightest resonances as explicit degrees of freedom is required.
\par
The resonance chiral effective theory with three flavours and only
including vector meson resonances is given, at the lowest chiral
order, by \cite{EG89}
\beq
{\cal L}_{\chi V} \, = \, {\cal L}_{\chi}^{(2)} \, + \,
{\cal L}_{KV} \, + \, {\cal L}_V^{(2)} \; .
\label{eq:lcr}
\eeq
Here ${\cal L}_{\chi}^{(2)}$ is the ${\cal O}(p^2)$ chiral lagrangian
\beq
{\cal L}_{\chi}^{(2)} \, = \, \Frac{F^2}{4} \, \langle \, u_{\mu}
u^{\mu} \, + \, \chi_+ \, \rangle \; ,
\label{eq:lc}
\eeq
where $F$ is the pion decay constant ($F \approx 92.4 \, \mbox{MeV}$),
\begin{eqnarray}
u_{\mu} \, =  \, i \, [ u^{\dagger} (\partial_{\mu} - i r_{\mu}) u
\, - \, u ( \partial_{\mu} - i \ell_{\mu}) u^{\dagger} \, ] \; , \; \; \;
\; \; \; \;   & \;  &
\; \; \; \; \; \; \chi_{+} \, = \, u^{\dagger} \chi u^{\dagger} \, + \,
u \chi^{\dagger} u \; ,  \nonumber
\\ \, \nonumber \\
u \,  =  \, \exp \left( \Frac{i}{\sqrt{2}F} \Pi \right) \; , \;
\; \; \; \;  \; \; \; \; \; \; \; \; \; \; \; \; \; \; \; \; \; \; \;
\; \; \; \; \; \; \; \; \; \; \; \; \;  & \, &
\label{eq:forma}
\end{eqnarray}
and $\Pi$ is the usual representation of the Goldstone fields
\beq
\Pi \, = \, \left( \begin{array}{ccc}
                   \Frac{\pi^0}{\sqrt{2}} \, + \,
           \Frac{\eta_8}{\sqrt{6}} & \pi^+ & K^+ \\
           \pi^- & - \Frac{\pi^0}{\sqrt{2}} \, + \,
           \Frac{\eta_8}{\sqrt{6}} & K^0 \\
           K^- & \overline{K^0} & - \Frac{2}{\sqrt{6}} \eta_8
           \end{array}
            \right) \; .
\label{eq:pi}
\eeq
In Eq.~(\ref{eq:lc}), $\langle A \rangle$ stands for a trace in
the flavour space.
${\cal L}_{\chi}^{(2)}$ has a $SU(3)_L \otimes SU(3)_R$ chiral gauge
symmetry supported by the external fields $\ell_{\mu}$, $r_{\mu}$
and $\chi$.
\par
In Eq.~(\ref{eq:lcr}), ${\cal L}_{KV}$ is the kinetic lagrangian of vector
mesons and ${\cal L}_V^{(2)}$ describes the chiral couplings of
vector mesons to the Goldstone fields and external currents at the
lowest order,
\beq
{\cal L}_V^{(2)} \, = \, \Frac{F_V}{2 \sqrt{2}} \, \langle V_{\mu \nu}
\, f_+^{\mu \nu} \, \rangle \; + \; i \, \Frac{G_V}{\sqrt{2}} \,
\langle \, V_{\mu \nu} \, u^{\mu} \, u^{\nu} \, \rangle \; ,
\label{eq:l2v}
\eeq
where $f_+^{\mu \nu} \, = \, u F_L^{\mu \nu} u^{\dagger} \, + \,
u^{\dagger} F_R^{\mu \nu} u$, with $F_{L,R}^{\mu \nu}$ the field
strength tensors of the left and right external currents $\ell_{\mu}$ and
$r_{\mu}$. $V_{\mu \nu}$
denotes the octet of the lightest vector mesons, in the antisymmetric
formulation \cite{GH84,EG89,KY69}, with a flavour content analogous to $\Pi$ in
Eq.~(\ref{eq:pi}). The effective couplings $F_V$ and $G_V$ can be
determined from the decays $\rho^0 \rightarrow e^+ e^-$ and
$\rho^0 \rightarrow \pi^+ \pi^-$ respectively.
Notice that only linear terms in the vector fields have been considered in
${\cal L}_V^{(2)}$.
\par
Assuming unsubtracted dispersion relations for the pion and axial
form factors one gets two constraints \cite{EG89a} among the couplings
$F$, $F_V$ and $G_V$ :
\begin{eqnarray}
F_V \, G_V \; = \; F^2 \; , \; \; & \; \; & \; \; \;
F_V \, = \, 2 \, G_V \; ,
\label{eq:const}
\end{eqnarray}
which are reasonably well satisfied phenomenologically.
We will enforce those constraints in our analysis.
\par
As thoroughly studied in Ref.\cite{EG89a} the use of the antisymmetric
formalism to describe spin--1 mesons simplifies the structure of the
effective action at the lowest chiral order in the even--intrinsic
parity sector.
If we use the Proca formulation of the vector
fields instead, we need to consider the ${\cal O}(p^4)$ chiral
lagrangian of Gasser and Leutwyler \cite{GH85}, with appropriate
$L_i$ coefficients, in order to fulfil the short--distance QCD
constraints. We will
comment later on about the independence of our results
on this particular choice.
\par
The bare propagator of the vector mesons,
in the antisymmetric formulation, is given by
\beq
\langle 0 | T \{ V_{\mu \nu}(x),V_{\rho \sigma}(y) \} | 0 \rangle \; =
\; \Frac{i}{M_V^2} \, \int \Frac{d^4q}{(2 \pi)^4} \,
\Frac{e^{-iq(x-y)}}{M_V^2 - q^2 - i\varepsilon} \, \left[
\, M_V^2 \, \Omega^L_{\mu \nu \rho \sigma} \, + \,
\Omega^T_{\mu \nu \rho \sigma} \, \right] \; ,
\label{eq:propa}
\eeq
with
\begin{eqnarray}
\Omega_{\mu \nu \rho \sigma}^L \, & \doteq & \,
g_{\mu \rho} g_{\nu \sigma} \, - \, g_{\mu \sigma} g_{\nu \rho} \; ,
\label{eq:omega} \\
\Omega_{\mu \nu \rho \sigma}^T \, & \doteq & \,
g_{\mu \rho} q_{\nu} q_{\sigma} \, - \, g_{\rho \nu} q_{\sigma} q_{\mu}
\, - \, q^2 g_{\mu \rho} g_{\nu \sigma}  \; - (\rho \leftrightarrow
\sigma) \; , \nonumber
\end{eqnarray}
that satisfies $q^{\mu} \Omega^T_{\mu \nu \rho \sigma} \, = \, 0$.
Chiral symmetry requires that the interaction between the vector mesons and
pseudoscalars or external currents is a derivative coupling, as shown in
${\cal L}_V^{(2)}$. Consequently, when the vector--meson
propagator connects with only one
external current or two pseudoscalars, the transverse part
$\Omega_{\mu \nu \rho \sigma}^T$ does not contribute.
\par
As commented above, a lagrangian density that includes the Goldstone
bosons and spin--1 resonances is not unique
but depends on the definitions of the fields, whereas the observable
physical quantities should be
independent of them. If we want to construct the dressed
propagator of the $\rho^0$ meson we should consider, for a definite 
intermediate state, all the
contributions carrying the appropriate quantum numbers. The first cut,
in the $\rho^0$ case, is a two--pseudoscalar absorptive 
contribution that happens to saturate its width. Here we will
take into account the two--particle absorptive contributions only;
higher multiplicity intermediate states being suppressed by phase
space and ordinary chiral counting.
The effective vertices, that will contribute to the observables we are
interested in, are those corresponding to an external vector current
coupled to two
pseudoscalar legs, and to a vector transition in the s--channel contributing
to the four pseudoscalar vertex. These transitions are not only
diagrammatically driven by the
$\rho$ propagator, but also through local contributions that have to be
included.
The construction of the effective vertices goes as sketched in Figure 1
where, at the leading order, the local vertices on the right--hand side
of the equivalence are provided by the ${\cal O}(p^2)$ chiral
lagrangian ${\cal L}_{\chi}^{(2)}$ in Eq.~(\ref{eq:lc}). The diagrams
contributing to physical observables will be constructed taking into
account all possible combinations of these two effective vertices.
\begin{figure}
    \begin{center}
       \setlength{\unitlength}{1truecm}
       \leavevmode
       \hbox{%
       \epsfysize=7.2cm
       \epsffile{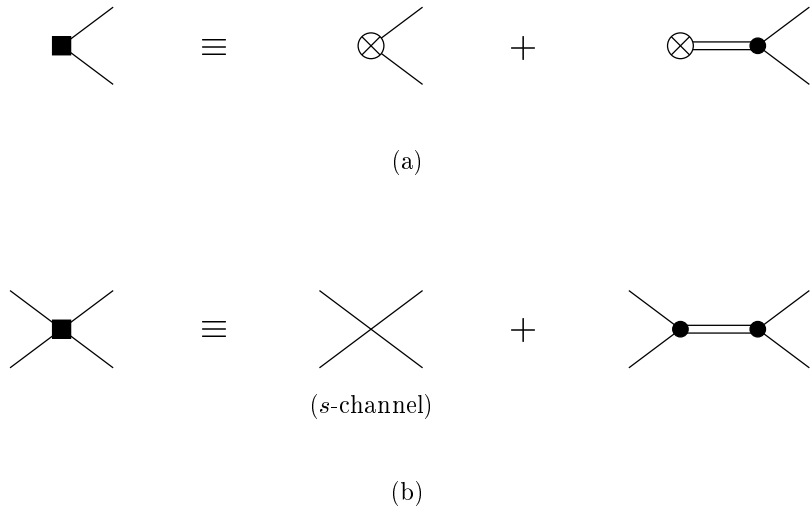}
       }
    \end{center}
    \caption{Effective vertices contributing to vector transitions
             in the s--channel that are relevant for the pion form
         factor and elastic $\pi \pi$ scattering. The crossed
         circle stands for an external vector current insertion. A
         double line indicates the $\rho^0$ meson and single ones
         the pseudoscalars. Local vertices on the right--hand side
         are provided, at leading order, by ${\cal L}_{\chi}^{(2)}$.}
    \protect\label{fig1}
\end{figure}
%
\par
Let us start with
the vector pion form factor $F_V(q^2)$ defined by
\begin{equation}
\langle \, \pi^+(p') \, | \, v_{\mu}^3 \, | \, \pi^+ (p) \, \rangle
\, = \,  \, (p+p')_{\mu} \, F_V(q^2) \; ,
\label{eq:fvdef}
\end{equation}
where $q^2 = (p-p')^2$ and $v_{\mu}^3$ is the third $SU(3)$ component
of the vector current $v_{\mu} = (\ell_{\mu} + r_{\mu})/2$. $F_V(q^2)$
is dominated by the contribution of the $\rho^0$ meson and has
thoroughly been studied in Ref.\cite{PT97} up to $E \sim 1.5 \gev$.
Clearly, one cannot describe $F_V(q^2)$ in that region of energy
using the bare $\rho^0$ propagator of Eq.~(\ref{eq:propa}); the
width of the resonance has to be
introduced to regulate the pole of the propagator.
\par
We propose then to construct a Dyson--Schwinger--like
equation through a perturbative loop expansion.
According to our previous discussion, at tree level
one has to take into account the amplitudes provided by Figures 2.a and 2.b,
i.e. the effective vertex in Figure 1.a.
The next step is to consider one--loop corrections. We are only interested
in those contributions with absorptive parts in the s--channel. They are
generated by
inserting a pseudoscalar loop using the two effective vertices in Figure 1
which leads
to the four contributions in Figures 2.c, 2.d, 2.e and 2.f. In this way
we have proceeded up to
two loops. The resulting infinite series happens to be geometric and
its resummation gives
\begin{equation}
F_V(q^2) \, = \, \Frac{M_V^2}{M_V^2 \, \left[ 1 \, + \,
2 \Frac{q^2}{F^2} \, \mbox{Re} \, \overline{B_{22}} \, \right] \,
- \, q^2 \, - \, i \, M_V \, \Gamma_{\rho}(q^2)} \; ,
\label{eq:fvres}
\end{equation}
where $M_V$ is the mass of the octet of vector mesons in the chiral
limit, $\overline{B_{22}} \equiv B_{22}[q^2,m_{\pi}^2,m_{\pi}^2] \, + \,
\frac{1}{2} B_{22}[q^2,m_K^2,m_K^2]$ and the $B_{22}[q^2,m^2,m^2]$
function is given in the Appendix. The width of the $\rho^0$ meson
$\Gamma_{\rho}(q^2)$ is given by
\beq
\Gamma_{\rho}(q^2) \, = \, - \, 2 \, M_V \, \Frac{q^2}{F^2} \, \mbox{Im}
\, \overline{B_{22}} \, = \,
\Frac{M_V \, q^2}{96 \, \pi \, F^2} \,
\left[ \, \sigma_{\pi}^3 \,
\theta(q^2 - 4 m_{\pi}^2) \, + \, \Frac{1}{2} \, \sigma_K^3\,
\theta(q^2 - 4 m_{K}^2)  \, \right] \; ,
\label{eq:rhooff}
\eeq
where $ \sigma_P \, = \, \sqrt{1-\Frac{4 m_P^2}{q^2}} $.
\par
\begin{figure}
    \begin{center}
       \setlength{\unitlength}{1truecm}
       \leavevmode
       \hbox{%
       \epsfysize=5.8cm
       \epsffile{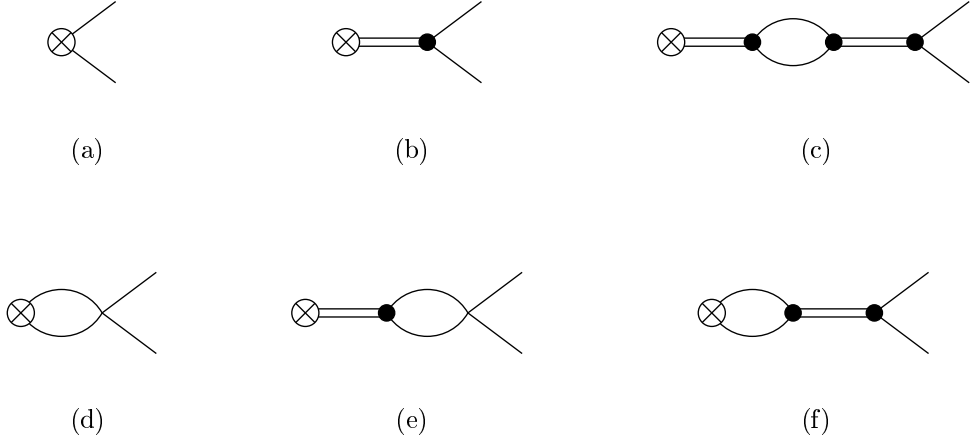}
       }
    \end{center}
    \caption{Diagrams contributing to the vector pion form factor
             up to one loop in the resonance chiral lagrangian
             given by ${\cal L}_{\chi V}$ that have an absorptive
             part in the s--channel. The lines and symbols stand for as
         in Figure 1.}
    \protect\label{fig2}
\end{figure}
%
The real part
of the pole of $F_V(q^2)$ in Eq.~(\ref{eq:fvres}) needs still to be
regulated through the wave function and mass renormalization of the vector
field. The local part of $\mbox{Re} B_{22}$ can be fixed in this case by
matching the result in Eq.~(\ref{eq:fvres}) with the well known expression
of $F_V(q^2)$ at one loop in $\chi$PT \cite{PT97,GH85b}. One gets then
\beq
\left. B_{22}[q^2,m_P^2,m_P^2]\, \right|_{F_V} \, = \, \Frac{1}{192 \, \pi^2} \,
\left[ \, \ln \left( \Frac{m_P^2}{M_V^2} \right) \, + \,
8 \, \Frac{m_P^2}{q^2} \, - \, \Frac{5}{3} \, + \, \sigma_P^3 \,
\ln \left( \Frac{\sigma_P + 1}{\sigma_P-1} \right) \, \right] \; .
\eeq
We have used the standard vector meson dominance value \cite{EG89} of
the relevant ${\cal O}(p^4)$
chiral coupling, $L_9(M_V^2) \, = \, \Frac{F^2}{2 M_V^2}$, which
agrees very well with its phenomenologically extracted value.
\par
To one--loop accuracy, Eq.~(\ref{eq:fvres}) agrees by construction
with the Omn\`es resummation of chiral logarithms performed in Ref.\cite{PT97}.
However, and as pointed out in Ref.\cite{GM91},
the Omn\`es resummation also reproduces the contribution of double
chiral logarithms, while our one--loop resummation does not \cite{FG98}.
This is
not surprising, since our result in Eq.~(\ref{eq:fvres}) has been obtained
by considering not all possible diagrams but only those which
are driven by the $\rho^0$ resonance and contribute to its width.
It would be interesting to perform a detailed study
of both resummations at the next chiral order and compare them with the
known two--loop $\chi$PT results \cite{BC98}.
\par
An analogous procedure can be applied to
the study of elastic $\pi \pi$ scattering. For definiteness we take the
$J=1$ transition in the s--channel amplitude of
$\pi^+ \pi^- \rightarrow \pi^+ \pi^-$.
This channel is again dominated by the $\rho^0$ meson
and we can proceed to construct a Dyson--Schwinger equation as in the case
of the vector pion form factor. Thus, we consider analogous diagrams to
those in Figure 2, substituting the external vector current insertions
by two pion legs, according to all the possible contributions of the
effective vertices in Figure 1.
By projecting the p--wave
(that corresponds to the $J=1$ contribution) we find again a geometric
series, which can be resummed to give
\beq
{\cal A}(\pi^+ \pi^- \rightarrow \pi^+ \pi^-) |_{J=1} \, = \,
\Frac{-i}{2 F^2} \, (\,u \, - \, t \,) \,
 \Frac{M_V^2}{M_V^2 \, \left[ 1 \, + \,
2 \Frac{q^2}{F^2} \, \mbox{Re} \, \overline{B_{22}} \, \right] \,
- \, q^2 \, - \, i \, M_V \, \Gamma_{\rho}(q^2)} \; ,
\label{eq:pipires}
\eeq
where $u$ and $t$ are the usual Mandelstam variables ($q^2 = s$). We see
that the pole
of the amplitude coincides with the one we got for the vector pion
form factor and, therefore, gives the same width of the $\rho^0$
meson. Contrary to the pion form factor case, here it is not possible
to fix the polynomial part of the $\mbox{Re} B_{22}$ function in a simple
way, because
a proper matching with the chiral low--energy expansion requires to take
into account p--wave contributions in the $t$ and $u$
channels, which are
not accounted for in our result of Eq.~(\ref{eq:pipires}).
\par
In the next section we will introduce a definition of the off--shell
width of spin--1 resonances that is consistent with our previous
results.

\section{The definition of a hadronic off--shell $\rho^0$ width}
\hspace*{0.5cm} We propose to define the spin--1 meson width
as the imaginary part
of the pole generated by resumming those diagrams, with an absorptive part
in the s--channel, that contribute to the two--point function of the
corresponding vector current. That is, the pole
of
\beq
\Pi_{\mu \nu}^{jk} \, = \, i \, \int \, d^4 x \, e^{iqx} \,
\langle 0 | \, T (V_{\mu}^j(x) V_{\nu}^k(0)) \, | 0 \rangle \; ,
\label{eq:two}
\eeq
with
\beq
V_{\mu}^j \, = \, \Frac{\delta S_{\chi V}}{\delta v_j^{\mu}} \; ,
\label{eq:curr}
\eeq
where $S_{\chi V}$ is the action associated to ${\cal L}_{\chi V}$.
\par
The $\rho^0$ quantum numbers correspond to $j=k=3$.
Lorentz covariance and current conservation
allow us to define an invariant
function of $q^2$ through
\begin{eqnarray}
\Pi_{\mu \nu}^{33} \, & = &  \, (q^2 g_{\mu \nu} \, - \, q_{\mu} q_{\nu}) \,
\Pi^{\rho}(q^2) \; , \nonumber \\ \, \label{eq:inva}  \\
\Pi^{\rho}(q^2) \, & = & \, \Pi_{(0)}^{\rho} \, + \, \Pi_{(1)}^{\rho} \, + \,
\Pi_{(2)}^{\rho} \, + \cdots  \; , \nonumber
\end{eqnarray}
where $\Pi_{(0)}^{\rho}$ corresponds to the tree level contribution of
Figure 3.a
, $\Pi_{(1)}^{\rho}$ to the one--loop amplitudes and so forth.
Up to one loop, and considering again the two--particle absorptive
contributions only, all the diagrams generated by the effective vertices
in Figure 1 are shown in Figure 3.
We find, in the isospin limit,
\beq
\Pi_{(0)}^{\rho} \, = \, \, \Frac{F_V^2}{M_V^2 - q^2} \; ,
\label{eq:tree}
\eeq
\beq
\Pi_{(1)}^{\rho} \, = \, \Pi_{(0)}^{\rho}
\, \left[ \, - \, \Frac{M_V^2}{F_V^2} \,
\Frac{M_V^2}{M_V^2 - q^2} \, 4 \, \overline{B_{22}} \, \right] \; .
\label{eq:1loop}
\eeq
\begin{figure}
    \begin{center}
       \setlength{\unitlength}{1truecm}
       \leavevmode
       \hbox{%
       \epsfysize=5.3  true cm
       \epsffile{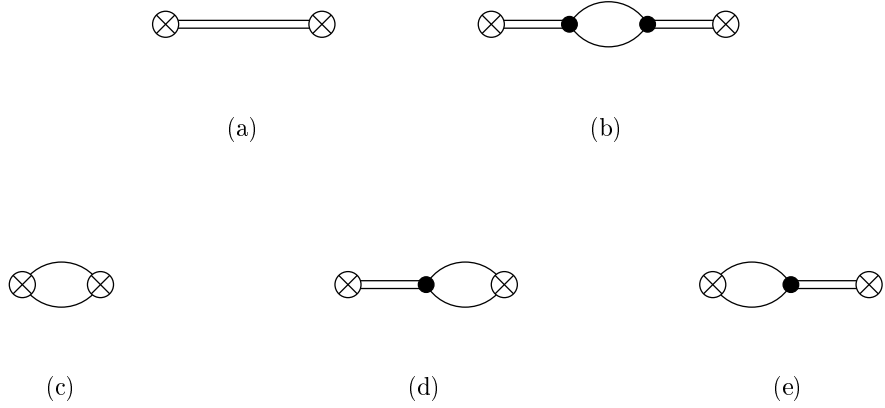}
       }
    \end{center}
    \caption{Diagrams contributing to the vector--vector correlator
    $\Pi_{\mu \nu}^{33}$ up to one loop. The lines and symbols stand for
    as in Figure 1.}
    \protect\label{fig3}
\end{figure}
%
\par
QCD predicts that two--point spectral functions of
vector currents go to a constant value as $q^2 \rightarrow
\infty$ \cite{FN79}. The loop diagram in Figure 3.b
behaves also as a constant for large values
of $q^2$, which is against the expectations because it corresponds
to only one of an infinite number of possible intermediate states.
In order to satisfy the QCD prediction, one would foresee that all
the individual
(positive) contributions from the different
intermediate states should vanish in the limit $q^2 \rightarrow \infty$.
Indeed, this is achieved in our case when diagrams in Figures 3.c, 3.d
and 3.e are added.
\par
Our result for $\Pi_{(1)}^{\rho}$ corresponds to a single one--loop
diagram with two effective vertices, as shown in Figure 4.a.
It vanishes
as $q^2 \rightarrow \infty$, as QCD requires, and this fact happens
at every higher order when all possible diagrams with absorptive
contributions in the s--channel (and not just
self--energies) are included.
\par
We proceed to evaluate $\Pi_{(2)}^{\rho}$, that is,
the contribution of two--loop diagrams that arise from
${\cal L}_{\chi V}$ with absorptive contributions in the s--channel.
These diagrams can be easily constructed by iterating in all
possible ways the one--loop diagrams shown in Figure 3;
they correspond to the single two--loop diagram
with three effective vertices shown in Figure 4.b. We obtain
\beq
\Pi_{(2)}^{\rho} \, = \, \Pi_{(1)}^{\rho} \,
\left[ \, - \, \Frac{q^2}{F_V^2} \, \Frac{M_V^2}{M_V^2 - q^2} \,
4 \, \overline{B_{22}} \, \right] \, \ \; .
\label{eq:2loop}
\eeq
A watchful look to the evaluation
(and a check up to three--loops) make us to conclude that the
invariant two--point function $\Pi^{\rho}(q^2)$, generated by resumming
effective loop diagrams with an absorptive amplitude in the s--channel
(as those in Figure 4), is perturbatively given by
\beq
\Pi^{\rho}(q^2) \, = \, \Pi_{(0)}^{\rho} \, + \,
\Pi_{(1)}^{\rho} \, \sum_{n=0}^{\infty}  \,
\left[ \, - \, \Frac{q^2}{F_V^2} \, \Frac{M_V^2}{M_V^2 - q^2} \,
4 \, \overline{B_{22}} \, \right]^n \; = \;
 \Pi_{(0)}^{\rho} \, \left[
\, 1 \, + \, \omega \, \sum_{n=0}^{\infty} \, \left( \, \Frac{q^2}{M_V^2} \,
\omega \, \right)^n\, \right] \; ,
\label{eq:pexp}
\eeq
where
\beq
\omega \, = \, - \, \Frac{M_V^2}{F_V^2} \, \Frac{M_V^2}{M_V^2 - q^2} \,
4 \, \overline{B_{22}} \; .
\label{eq:omiga}
\eeq
\begin{figure}
    \begin{center}
       \setlength{\unitlength}{1truecm}
       \leavevmode
       \hbox{%
       \epsfysize=2.3  true cm
       \epsffile{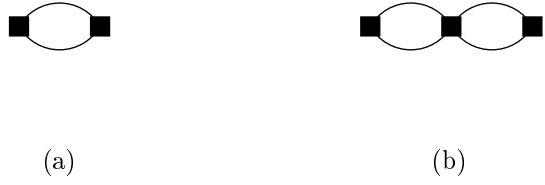}
       }
    \end{center}
    \caption{One and two--loop diagrams leading to $\Pi_{(1)}^{\rho}$ (a)
    and $\Pi_{(2)}^{\rho}$ (b). The
    effective squared vertices are given in Figure~1.}
    \protect\label{fig4}
\end{figure}
%
Now, resumming, using that $F_V^2 = 2 F^2$ (see Eq.~(\ref{eq:const})), and
substituting the expression of $\omega$, we finally get
\beq
\Pi^{\rho}(q^2) \, = \,
 \, \Frac{2 \, F^2}{M_V^2 \, \left[ 1 \, + \,
2 \Frac{q^2}{F^2} \, \mbox{Re} \, \overline{B_{22}} \, \right] \,
- \, q^2 \, - \, i \, M_V \, \Gamma_{\rho}(q^2)} \,
\left[ \, 1 \, - \, 2 \Frac{M_V^2}{F^2} \, \overline{B_{22}}
\, \right] \; ,
\label{eq:final}
\eeq
where the off--shell $\rho^0$ width $\Gamma_{\rho}(q^2)$ is given
by Eq.~(\ref{eq:rhooff}).
\par
We emphasize that our only concern with the result of $\Pi^{\rho}(q^2)$
in Eq.~(\ref{eq:final}) is the reconstruction of the imaginary part
of the pole. Our evaluation of the two--point function vector--vector
correlator is far from complete, since we have considered only those
diagrams with absorptive contributions in the
s--channel\footnote{In fact, those diagrams where the vector fields
are not explicit (such as that in Fig.~3.c) give rise to a
dispersive divergent piece proportional
to $q_{\mu} q_{\nu}$ that we disregard. It is also clear from
our result in Eq.~(\ref{eq:final}) that the real part of the pole
needs regularization.}. The only significative result of $\Pi^{\rho}(q^2)$
is its imaginary part. The residue in $\Pi^{\rho}(q^2)$
deserves a further comment. While
it carries an imaginary piece, this result is proper as far as it
satisfies the required unitarity condition
\beq
\mbox{Im}\,\Pi^{\rho}(q^2) \, = \, \Frac{1}{48 \pi} \, \left[ \,
\sigma_{\pi}^3 \, \theta (q^2 - 4 m_{\pi}^2) \,  + \, \Frac{1}{2} \,
\sigma_K^3 \, \theta (q^2 - 4 m_K^2) \, \right] \,
| \, F_V(q^2) \, |^2 \; ,
\label{eq:unitari}
\eeq
with $F_V(q^2)$ given by our result in Eq.~(\ref{eq:fvres}). This
shows the consistency of our resummation procedure.

\section{Independence of the spin--1 field definition}
\hspace*{0.5cm}
In Ref.\cite{EG89a} it was shown that, at ${\cal O}(p^4)$ in the chiral
expansion, sensible redefinitions of the spin--1 fields give
the same results for physical low--energy observables. The equivalence between
redefinitions may require the presence of local terms that have
to be added in order to satisfy the short--distance QCD constraints.
In that reference it was concluded that, at
${\cal O}(p^4)$, the antisymmetric formulation for spin--1 meson
resonances is the only one that does not require the presence of
such local contributions and is, therefore, the simplest one.
\par
However, a resummation of the two--point vector--vector correlator
when vectors are active degrees of freedom has no defined chiral
counting. Therefore, one might worry that $\Gamma_{\rho}(q^2)$ could depend
on the field formulation used to describe the vector mesons.
\par
In order to see the independence of our results on the spin--1 field
formulation, it is enough to realize that the effective vertices defined
in Figure 1 are universal. Different theoretical descriptions of the
vector (or axial--vector) meson degrees of freedom lead to resonance--exchange
contributions which differ by local terms. Since the physical amplitudes
are constrained to satisfy the appropriate QCD behaviour at large momenta,
this difference is necessarily compensated by explicit local terms
(with fixed couplings) \cite{EG89a}. Including those local terms
in the local vertices of Figure 1, the resulting effective vertices
are formulation independent.
\par
Since our resummation has been based on these effective vertices, the
universality of our result follows.
\par
Our discussion has been carried out for a two--particle cut, which is
the most important contribution
for the $\rho^0$ and other vector mesons. A similar procedure could be
applied to higher multiplicity intermediate states by constructing
the relevant effective vertices analogous to those in Figure 1. Although
technically much more involved, its study would be necessary to evaluate
the width of other resonances such as the $\omega$(872) or the 
axial--vector mesons.

\section{Conclusions}
\hspace*{0.5cm} In this work we have studied the off--shell width
of spin--1 meson resonances in a model--independent framework provided
by the resonance chiral effective theory of QCD at low energies.
We have performed resummations for the vector pion form factor and
the $\pi \pi \rightarrow \pi \pi$ ($J=1$) amplitude, showing that
they provide
the same structure for the pole of the $\rho^0$ vector meson.
In both cases the resummations correspond to geometric
Dyson--Schwinger--like series that include only diagrams with
absorptive contributions in the s--channel.
\par
We have defined the width of spin--1 meson resonances as given by the
imaginary part of the pole generated by resumming
loop diagrams, in the two--point correlator of vector or axial--vector currents
of the resonance chiral lagrangian, that have absorptive contributions in
the s--channel.
The width generated in this way satisfies the requirements
of analyticity, unitarity, chiral symmetry and the correct
asymptotic behaviour as prescribed by QCD.
\par
We have applied this procedure to evaluate the off--shell width
of the $\rho^0$ meson and we have worked out in detail the result.
Moreover we have shown that this definition
is independent of the formulation employed to describe the vector
meson fields. Hence our procedure can be applied straightaway to
evaluate the widths of other spin--1 resonances.
\vspace*{0.5cm} \\
\noindent {\large \bf Acknowledgements} \\ \\
\hspace*{0.5cm}
This work has been supported in part by TMR, EC--Contract No.\
ERB FMRX-CT98-0169 and by CICYT (Spain) under grants PB97-1261
and AEN99-0692. D.G.D.\ acknowledges the warm hospitality given
by the Departament
de F\'{\i}sica Te\`orica of the University of Valencia, and the
financial support provided by Fundaci\'on Antorchas, Argentina.
J.P.\ wishes to thank F.J.\ Botella for interesting discussions
on the subject of this paper.

\appendix
\newcounter{homero}
\renewcommand{\thesection}{\Alph{homero}}
\renewcommand{\theequation}{\Alph{homero}.\arabic{equation}}
\setcounter{homero}{1}
\setcounter{equation}{0}

\section*{Appendix}
\hspace*{0.5cm} The function $B_{22}[q^2,m^2,m^2]$ used in the text
is defined through
\beq
\int \, \Frac{d^D \ell}{i \, (2 \pi)^D} \,
\Frac{\ell_{\mu} \ell_{\nu}}{[\ell^2 - m^2] [(\ell - q)^2 - m^2]} \,
\equiv \,  q_{\mu} q_{\nu} \, B_{21} \, + \, q^2 g_{\mu \nu}
\, B_{22}  \; ,
\eeq
as
\begin{eqnarray}
B_{22}[q^2,m^2,m^2] \, & = & \, \Frac{1}{192 \, \pi^2} \,
\left[ \, \left( 1 - 6 \Frac{m^2}{q^2} \right) \,
\left[ \lambda_{\infty} \, + \,
\ln \left( \Frac{m^2}{\mu^2} \right) \, \right] \,
   + \, 8 \, \Frac{m^2}{q^2} \, - \, \Frac{5}{3} \, \right. \nonumber \\
 & &  \\
 & & \left. \; \; \; \; \; \; \; \; \; \; \; \; \; \; \;
+  \, \sigma^3 \,
 \ln \left( \Frac{\sigma + 1}{\sigma-1} \right)
\, \right] \; , \nonumber
\label{eq:q2b22}
\end{eqnarray}
where $\sigma = \sqrt{1 - 4 m^2/q^2}$ and
$\lambda_{\infty} = \frac{2}{D-4} \, \mu^{D-4} \,  -
( \Gamma'(1) + \ln (4 \pi) + 1)$.

\end{document}